\documentclass[11pt]{article}
\usepackage{moriond,epsfig}

\def\be{\begin{equation}}
\def\ee{\end{equation}}
\def\bea{\begin{eqnarray}}
\def\eea{\end{eqnarray}}

\newcommand{\Npix}{N_{\rm pix}}

\begin{document}
\vspace*{4cm}
\title{CMB Cosmological Parameter Estimation: Methods and Current Results}

\author{ \textbf{ M.~Douspis}, J.G.~Bartlett,  A.~Blanchard \& M.~Le~Dour}

\address{Observatoire Midi-Pyr\'en\'ees, 31400 TOULOUSE}

\maketitle\abstracts{The majority of present efforts to constrain cosmological
parameters with cosmic microwave background (CMB) anisotropy data
employ approximate likelihood functions, the time consuming
nature of a complete analysis being a major obstacle.
We have performed a full (unapproximated)
likelihood analysis on several experiments that allows us to 
examine the various assumptions made in these approximate 
methods and to evaluate their performance.  Our results indicate that
care should be taken when using such approaches.  With an improved
approximate method, we present some constraints on cosmological parameters 
using the entire present CMB data set.}

\section{Introduction}
        Since their first detection by COBE, the CMB temperature fluctuations 
have become an essential tool for constraining cosmological parameters. 
The number of experiments measuring anisotropies at different scales is now 
close to 20. With time, the observations become more accurate and their angular
resolution improves, increasing the number of pixels. This evolution towards 
higher precision measurements of CMB fluctuations inhibits complete 
(unapproximated) likelihood analyses over large sets of models 
(or free parameters)\footnote{note that the computation time grows like $\Npix ^3$}. 
Approximate methods have thus been proposed in order to explore as
much parameter space as possible with the complete CMB data set. 
Easy to use and fast, $\chi^2$ minimisation has been widely applied.
The approximate nature of this method resides in two assumptions: (i) the
one dimensional bandpower distribution is Gaussian; 
(ii) all pertinant information in a map is contained in the flat--band power. 
The latter is also assumed by other approximations, such as proposed
by Bond et al. (1998), Wandelt et al. (1998) and Bartlett et al. (1999).
Focussing on a small number of experiments (COBE, SASKATOON and MAX), we 
have performed a complete likelihood analysis over a 
reasonably representative set of (Inflationary) models.  This allows us 
to compare our results with those from a $\chi^2$
minimisation and to quantitatively examine these two assumptions. 

\section{Testing the assumptions}

In an inflationary scenario, the sky pixels are Gaussian distributed,
and so also the $a_{lm}'s$ in a spherical harmonic decomposition. 
But what is used in $\chi^2$ minimisation is the in--band power of 
the fluctuations; this corresponds to the square root of the 
temperature fluctuation variance and is not a 
Gaussian distributed quantity. This motivated us to look 
more carefully at the first assumption.  
Figure 1 shows an example where a 2-tailed Gaussian (with asymmetric
errors; dot--dot--dashed line) is a rather bad approximation to the true
likelihood curve (solid line). Even if the Gaussian appears relatively 
faithful near the maximum, it rapidly deviates from the likelihood function 
with the distance from the peak: the $\chi^2$ is thus very (overly) sensitive 
to the presence of outliers (which will be more common than it
expects). By comparison, the approximation developed in Bartlett et al.
 (1999)  
reproduces quite well the likelihood function\footnote{Many examples of the 
comparison between a Gaussian, the likelihood function and our
approximation may be seen at http://webast.ast.obs-mip.fr/cosmo/CMB}.
More accurate (relative to a full likelihood treatment) parameter
estimation can thus be obtained by using such
approximate likelihood forms, or simply by direct interpolation
of the one--dimensional in--band likelihood function,
when available.
 
\begin{figure}
\begin{center}
\psfig{figure=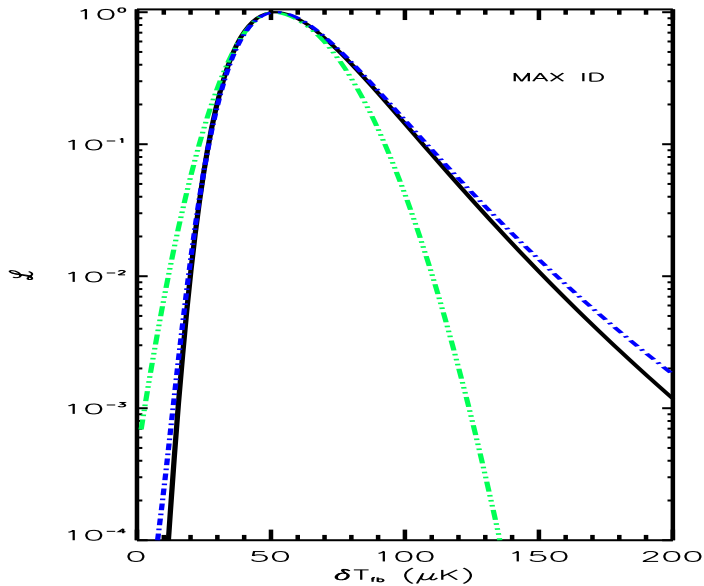,height=8cm, width=10cm}
\end{center}
\caption{True likelihood (solid line), 2-tailed Gaussian (dot--dot--dashed)
 and our approximation (Bartlett et al. 1999, dot--dashed) for the MAX
 ID field.
\label{fig:radish}}
\end{figure}

The second assumption supposes that the flat--band power estimate contains 
all the information in a map.  To check this assumption, we 
computed the flat--band power and its error bars for several 
bins of MAX and Saskatoon and will now compare the constraints
suggested by these estimates to those from a full likelihood analysis.  
This is illustrated in Figure 2. 
By definition, models passing within the error bars are considered
consistent with the flat--band estimates at the ``one sigma confidence level''.
This may be compared to the confidence level assigned by the
full likelihood analysis: focussing on just the MAX PH point 
(at the extreme left in Figure 2--left), the likelihood analysis excludes
the model plotted as the dotted line at more than 68\%, but 
accepts the model shown as the dashed line, all in flagrent 
contradiction with the flat--band estimate.  This can be understood
by looking at Figure 2--right.  Here, we show the results of a 
likelihood analysis over a family of spectra modeled by their
in--band power ($\delta T_N$) and slope ($m$): 
$\delta T(\ell)=\delta T_N. (\ell/\ell_{eff})^m$. 
Figure 2--right displays likelihood contours in the plane
 ($\delta T_N, m$) for MAX PH. 
We clearly see that MAX PH prefers spectra with positive slope and 
lower in--band power than the flat--band estimate (corresponding
to the vertical line at m=0).  The latter considers only the information 
available along the vertical line in Figure 2--right and
looses signal contained in the rest of the plane.
The symbols position the two models plotted in Figure 2--left, explaining
the disagreement between the two analysis methods.
The conclusion is that, at least in this example, the information contained
in the original pixel set cannot be reduced to a flat--band estimate. 
In the case of MAX PH, any approximation using only 
the flat--band power will never result in the same constraints as 
a full likelihood analysis. 

	Incorporating all the information available in a map 
(in--band power, slope, ...)\footnote{Depending on the signal--to--noise,
one could imagine the need for additional parameters.}
  will give more realistic constraints than possible with just
flat--band estimates.  A simple and efficient way to impliment this
is by interpolation the likelihood {\em surface} over $\delta T_N$ and $m$
for each bin (instead of the {\em one--dimensional} curve for the flat--band power).

\begin{figure}
\psfig{figure=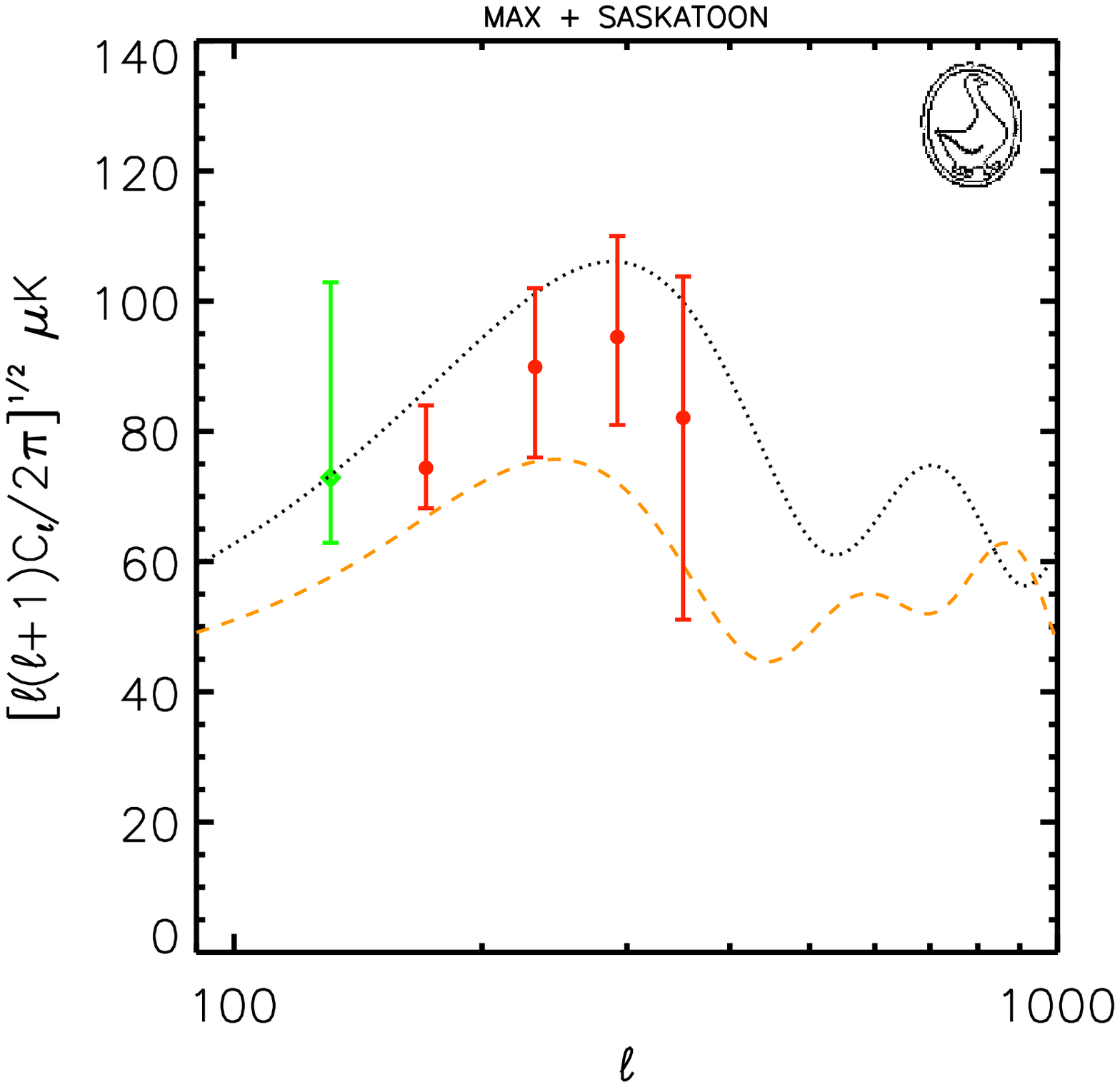,height=8cm}
\psfig{figure=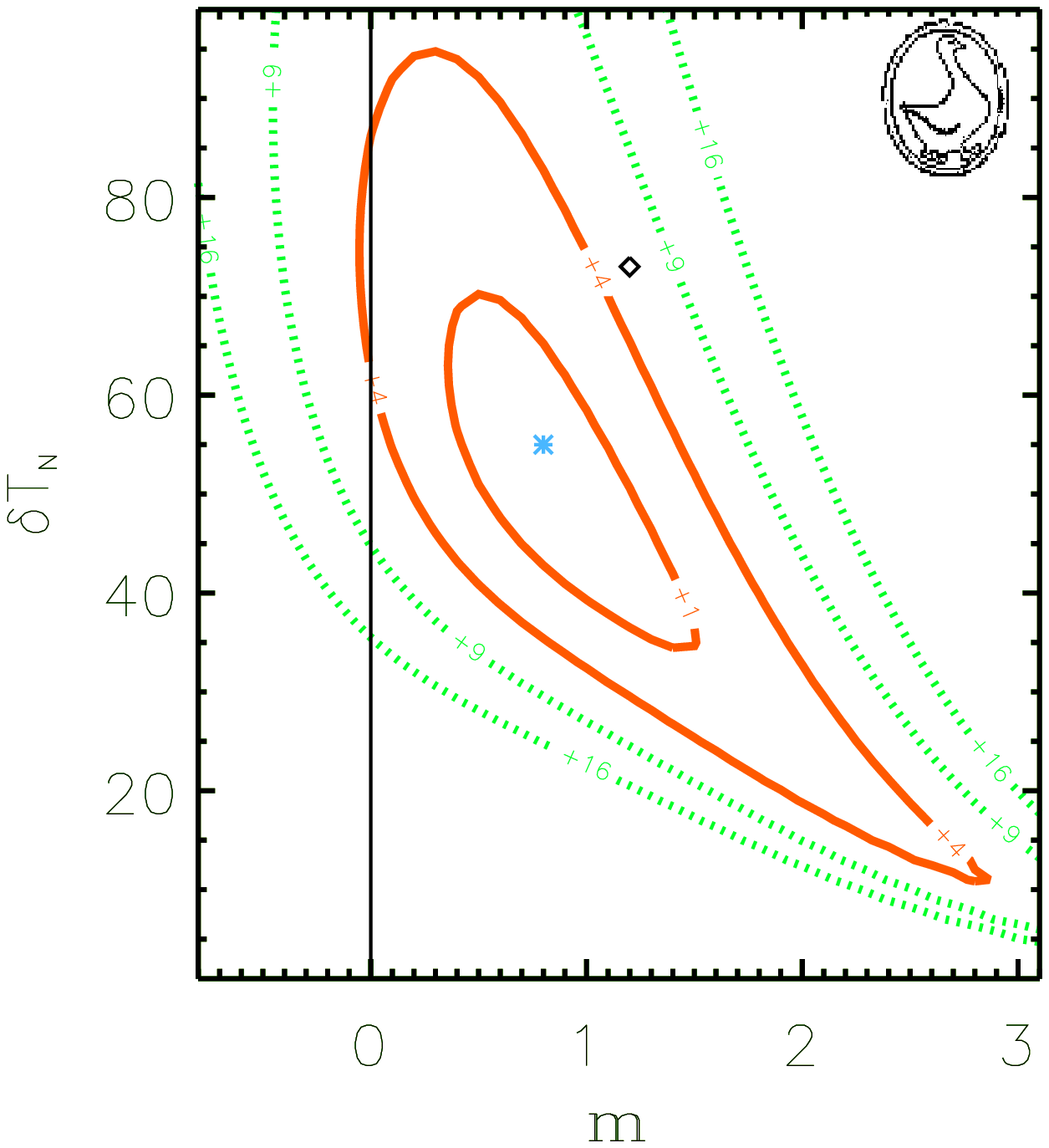,height=8cm} 
\caption{\textbf{LEFT}: Flat--band estimates for MAX PH and 4 bins of
Saskatoon CAP. \textbf{RIGHT}: Likelihood contours over spectral power versus
slope for MAX PH (see text).  Flat--band power estimates are restricted
to the vertical line.  
\label{fig:radish}}
\end{figure}


\section{Current Results}

We took into account the above remarks in an attempt to obtain
a more complete parameter estimation using the entire CMB data set
and a large set of Inflationary models.  In practice,
we used our approximation for those experiments for which only
the flat--band power estimate was given; interpolated the flat--band
likelihood curve when it was available; and interpolated the 
likelihood {\em surfaces} (power, slope) of Saskatoon and MAX. 
Our analysis corresponds to 70 different data points representing 20 experiments,
and more than 10 million models.  The compute time required was
350 cpu--hours on a DEC workstation -- equivalent to a simple $\chi^2$ 
minimisation\footnote{For comparison, the estimated time for a full
likelihood treatment with the same data set 
and models would be $\sim 10^5$ cpu--hours}. 

	Figures 3 show our principle results in the framework of
open Inflationary models with a cosmological constant.  The constraints are
given as 2--dimensional (approximated) likelihood contours at roughly 
$68$ and $95\%$ confidence.  In each case, the other investigated parameters 
$(H_0, Q, n, \eta_{10})$ are marginalised by projection.
A large number of the models are excluded in the plane $(\Omega_k, \lambda_0)$.
Prefered models have very low curvature, although no the cosmological
constant is relatively unconstrained.  
Figure 3--right is a nice summary of the status of the simplest of 
Inflationary 
predictions, namely that $n\sim 1$ and $\Omega_k=0$.  These values are
 indeed favored 
by current CMB observations.


\begin{figure}
\psfig{figure=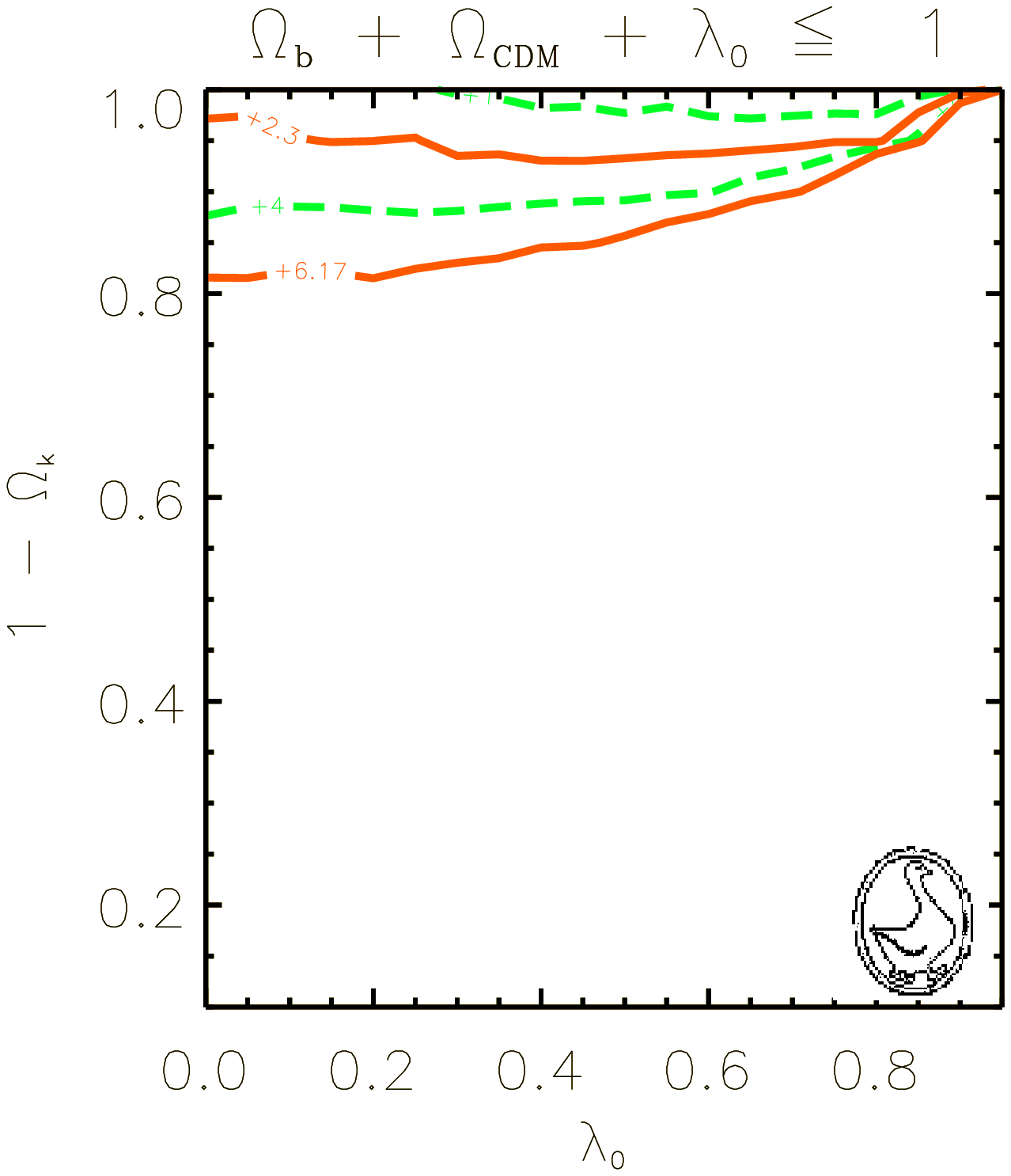,height=8cm}
\psfig{figure=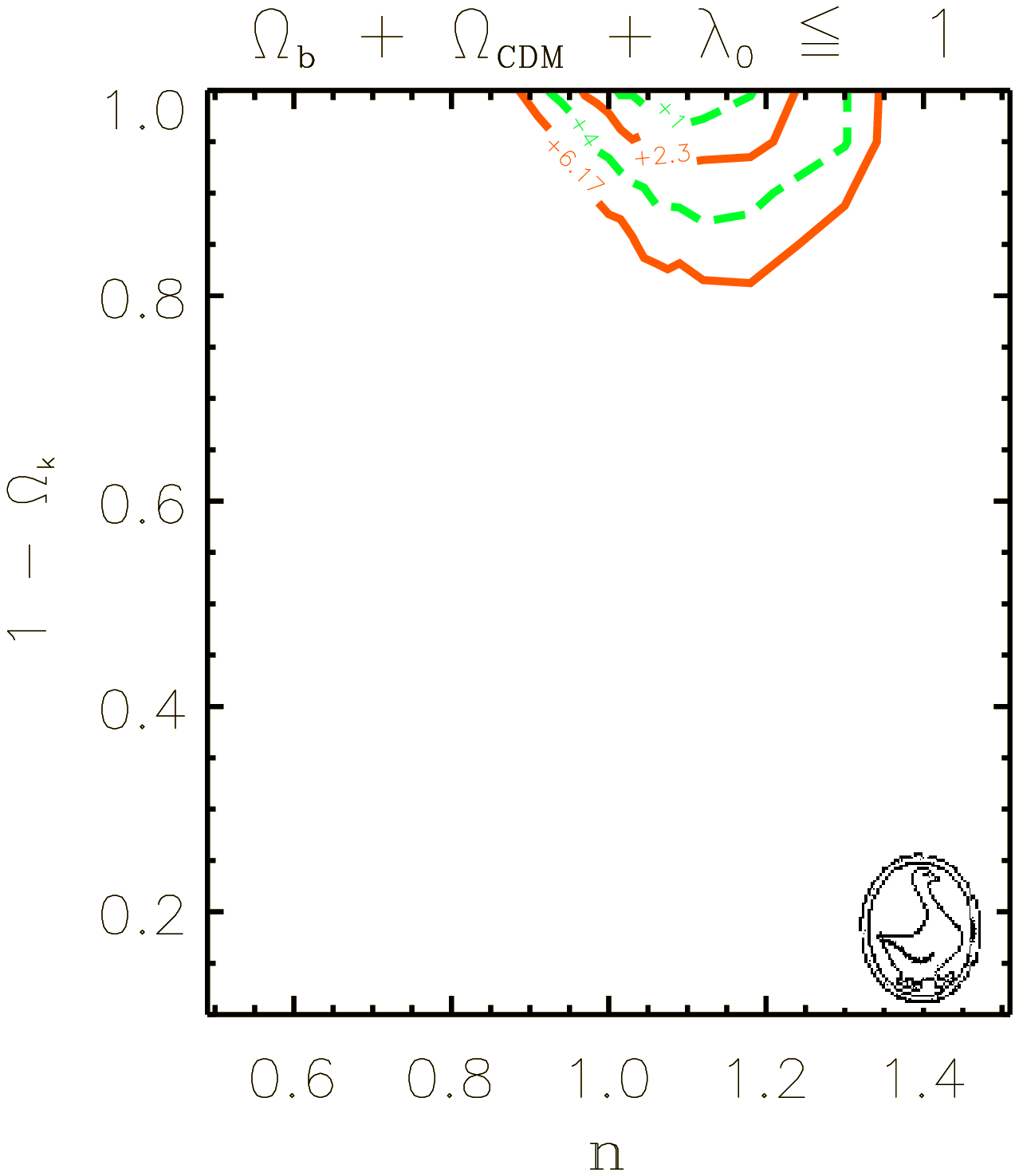,height=8cm}
\caption{\textbf{LEFT}: Constraints in ($\Omega_{tot}, \lambda_0$) plane 
obtained with our 
method. \textbf{RIGHT}: Constraints in in the ($\Omega_{tot}, n$) plane.
\label{fig:radish}}
\end{figure}

\section{Conclusion}
	
	We have seen how in some cases a Gaussian is a bad representation of
the one--dimensional in--band power likelihood function. 
To improve on $\chi^2$ methods, we therefore recommend the use 
of other, more appropriate approximations, as proposed in the literature,
or direct interpolation of the exact likelihood curve, when available.
Even this may not be entirely satisfactory because flat--band power estimates 
do not always sufficiently represent a CMB data set.  We 
have demonstrated that some of MAX and Saskatoon data (bins) prefer 
non--zero spectral slopes and different in--band powers.  In these
cases, the incorporation of additional information, for example,
the in--band spectral slope, should lead to a better 
reproduction of the actual constraints.  
Taking these remarques into account, have also seen how the first generation 
experiments favor the simple Inflationary scenario predictions: 
curvature close to zero and spectral index close to 
one.





\section*{References}
\begin{thebibliography}{99}

\bibitem{bjk} Bond J.R., Jaffe A.H. \& Knox L. 1998, astro--ph/9808264

\bibitem{us99} Bartlett J.G., Douspis M., Blanchard A. \& Le Dour M. 1999, astro--ph/9903045


\bibitem{us00} Douspis M., Bartlett J.G.,  Blanchard A. \& Le Dour M. 2000, in preparation 

\bibitem{us00} Wandelt B. J., Hivon E. \& Gorski K.M. 1998, astro--ph/9808292 


\end{thebibliography}
\end{document}